\documentclass[3p,times,procedia]{elsarticle}
\usepackage{nupha_ecrc}

\volume{00}

\firstpage{1}

\journalname{Nuclear Physics A}

\runauth{}

\jid{nupha}

\jnltitlelogo{Nuclear Physics A}

\usepackage{amssymb}

\usepackage[figuresright]{rotating}

\begin{document}

\begin{frontmatter}



\dochead{XXVIIIth International Conference on Ultrarelativistic Nucleus-Nucleus Collisions\\ (Quark Matter 2019)}

\title{Connecting far-from-equilibrium hydrodynamics to resummed transport coefficients and attractors}

\author[1]{Gabriel S. Denicol} 
\address[1]{Instituto de F\'isica, Universidade Federal Fluminense, UFF, Niter\'oi, 24210-346, RJ, Brazil}

\author[2]{Jorge Noronha} 
\address[2]{Department of Physics, University of Illinois, 1110 W. Green St., Urbana IL 61801-3080, USA}

\begin{abstract}
We investigate whether hydrodynamic attractors are present in simulations of the quark-gluon plasma formed in heavy-ion collisions. We argue that Lagrangian schemes to solve the relativistic viscous fluid equations can be particularly useful to characterize the properties of attractors in heavy ion collisions. Preliminary results are shown in Israel-Stewart theory to support such a claim. We also discuss how to perform the slow-roll expansion in Israel-Stewart theory undergoing a general flow, which naturally leads to the definition of a resummed shear viscosity coefficient that depends on the gradients of the hydrodynamic fields. 
\end{abstract}


\end{frontmatter}


\section{Introduction}
\label{intro}

Fluid dynamics is usually understood via an expansion around local equilibrium. The small parameter in such an expansion is codified by the Knudsen number, $K_N \sim \ell/L$, which denotes the ratio between the relevant microscopic/transport length scale, $\ell$, and a scale, $L$, associated with the gradients of conserved quantities. The general understanding is that fluid dynamics emerges in the regime where $K_N \ll 1$. However, in heavy ion collisions the Knudsen number is generally not small \cite{Niemi:2014wta,Noronha-Hostler:2015coa} and the situation only gets worse in small systems (such as in pA or pp collisions). Yet, hydrodynamics still seems to make sense even under those extreme conditions (see, e.g. \cite{PHENIX:2018lia}). This apparent paradox must be resolved in order to understand the (surprising) effectiveness of fluid dynamic approaches in heavy ion collisions. 

What defines the regime of applicability of hydrodynamics? The current working hypothesis in the field is that hydrodynamics may be defined as a universal attractor \cite{Heller:2015dha} where the dissipative currents display universal behavior that is independent of their initial conditions. In this contribution we investigate whether hydrodynamic attractors are present in simulations of the quark-gluon plasma formed in heavy ion collisions. We show preliminary results, computed using a Lagrangian solver, that support such a claim. We also show that the slow-roll expansion \cite{Heller:2015dha} in Israel-Stewart theory can be used to define a resummed shear viscosity coefficient that carries information about the underlying state of the fluid via a nontrivial dependence on the Knudsen number. 

\section{Are there hydrodynamic attractors in heavy ion collisions?}
\label{sec2}

Attractor behavior is well understood in simple systems under highly symmetrical flow conditions such as in Bjorken flow \cite{Heller:2015dha} (see \cite{Denicol:2017lxn,Denicol:2019lio} for analytical results). An important exception is Ref.\ \cite{Romatschke:2017acs}, which investigated the problem of attractor dynamics in spatially inhomogeneous systems. Indication of attractor behavior was found numerically in flow configurations similar to those experienced by the QGP in heavy ion collisions. However, despite this important step forward, a more general formulation of far-from-equilibrium fluid dynamics and attractor behavior, valid under general flow conditions, is still needed to correctly assess their role in the phenomenology of heavy ion collisions.

To further identify and characterize the properties of attractors in realistic simulations, we propose to think about this problem in a new way. The attractor was easily found in Bjorken flow because in this (transversely) homogeneous case the flow velocity is exactly $u^\mu = (1,0,0,0)$ (in appropriate coordinates \cite{Denicol:2014tha}) for all fluid cells. In terms of a Lagrangian description of the hydrodynamic evolution \cite{rezzolla}, in this Bjorken case all Lagrangian fluid particles are equivalent and, thus, they share the same attractor. Consider now the inhomogeneous case where the viscous hydrodynamic evolution is solved using a Lagrangian solver, such as the relativistic Smoothed Particle Hydrodynamics (SPH) algorithm \cite{Aguiar:2000hw,Denicol:2009am}, as in the v-USPhydro code \cite{Noronha-Hostler:2015coa,Noronha-Hostler:2013gga,Noronha-Hostler:2014dqa}. In Lagrangian schemes there is no fixed grid (in contrast to Eulerian methods) and the motion of the expanding fluid is described  in terms of SPH fluid elements whose individual trajectories are tracked over (proper)time. If there is an attractor, the dissipative currents (such as the shear stress tensor) for each SPH particle should approach a nontrivial universal regime for different initial conditions. However, due to the transverse expansion, each fluid element should have its own attractor (in contrast, in Bjorken flow all fluid elements are the same).

In this work, we investigated the presence of attractors in conformal \cite{Baier:2007ix,Marrochio:2013wla} Israel-Stewart (IS) theory \cite{Israel:1979wp} (at zero chemical potential).  The dynamics is given in terms of the variables $\{\varepsilon,u_\mu,\pi_{\mu\nu}\}$, where $\varepsilon$ is the local energy density, $u_\mu$ (with $u_\mu u^\mu = 1$) is the 4-velocity of the medium, and $\pi_{\mu\nu}$ is the shear stress tensor (which obeys the constraints $\pi^\mu_\mu=0$ and $u_\mu \pi^{\mu\nu}=0$). The equations of motion for these variables are defined by energy and momentum conservation, i.e. $\nabla_\mu T^{\mu\nu}=0$, with the energy-momentum tensor given by $T_{\mu\nu} = \varepsilon u_\mu u_\nu - P\Delta_{\mu\nu}+\pi_{\mu\nu}$ and $\Delta_{\mu\nu} = g_{\mu\nu}-u_\mu u_\nu$ ($g_{\mu\nu}$ is the spacetime metric), which must be solved together with the following additional equation for the shear stress tensor
\begin{equation}
\tau_\pi \left(D\pi^{\langle \mu\nu\rangle} + \frac{4}{3}\theta \pi^{\mu\nu} \right)+\pi^{\mu\nu} = 2\eta \sigma^{\mu\nu}.
\label{eq1}
\end{equation} 
Above, $\tau_\pi$ is the shear relaxation time, $D= u^\mu \nabla_\mu$, $\theta = \nabla_\mu u^\mu$, $\eta$ is the shear viscosity, $A^{\langle \mu\nu\rangle} = \Delta^{\mu\nu}_{\alpha\beta}A^{\alpha\beta}$ with $\Delta^{\mu\nu}_{\alpha\beta} = 1/2\left(\Delta^\mu_\alpha \Delta^\nu_\beta +\Delta^\mu_\beta \Delta^\nu_\alpha \right)-\Delta_{\alpha\beta}\Delta^{\mu\nu}/3$, and $\sigma^{\mu\nu} = \Delta^{\mu\nu}_{\alpha\beta}\nabla^\alpha u^\beta$ is the shear tensor. 

\begin{figure}
\centering
\includegraphics[width=0.9\textwidth]{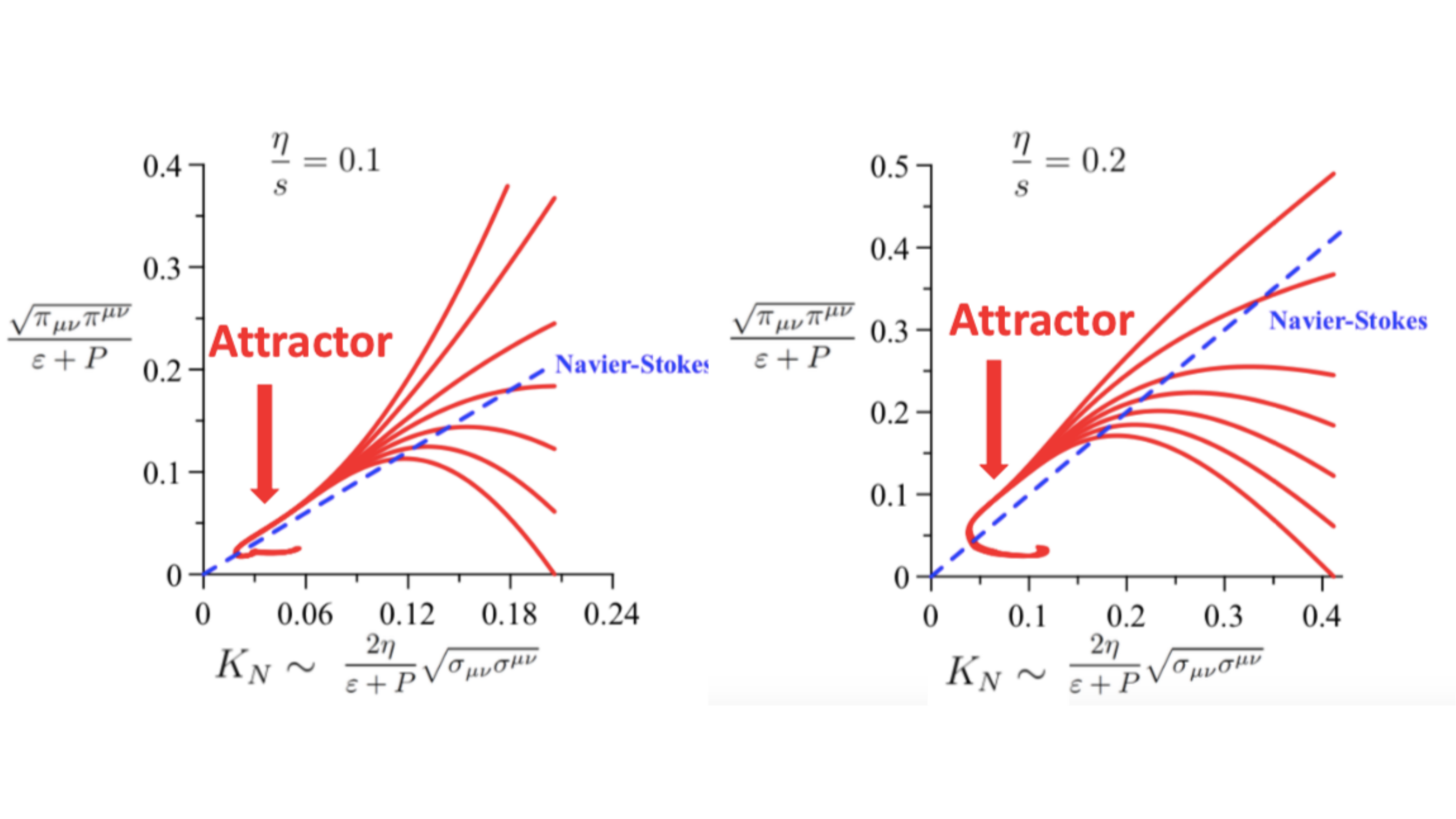}
\caption{Dependence of $\sqrt{\pi_{\mu\nu}\pi^{\mu\nu}}/(\varepsilon+P)$ on $2\eta \sqrt{\sigma_{\mu\nu}\sigma^{\mu\nu}}/(\varepsilon+P)$ for a fluid element at the center of the fireball. In the left panel $\eta/s=0.1$ while for the right panel $\eta/s=0.2$.}  
\label{fig1}
\end{figure}

In this conformal setup, the pressure $P=\varepsilon/3$ (for simplicity, a noninteracting gas of massless quarks and gluons is used), $\eta/s$ is constant (where $s$ is the entropy density) and we assume $\tau_\pi = 5\eta/(sT)$ with $T \sim \varepsilon^{1/4}$ being the temperature. In this contribution, these equations are solved using the SPH algorithm \cite{Denicol:2009am} assuming longitudinal boost invariance and radial symmetry in the transverse plane, with initial conditions corresponding to central smooth Glauber Au+Au collisions. We show in Fig.\ \ref{fig1} the behavior of $\sqrt{\pi_{\mu\nu}\pi^{\mu\nu}}/(\varepsilon+P)$ as a function of $2\eta \sqrt{\sigma_{\mu\nu}\sigma^{\mu\nu}}/(\varepsilon+P)$ for a fluid element at the center of medium. The maximum temperature at the center is 400 MeV and $\eta/s=0.1$ (left panel). Different initial conditions for $\pi^{\mu\nu}$ correspond to the solid red lines while the dashed blue line defines the Navier-Stokes (NS) regime where $\pi^{\mu\nu} = 2\eta\sigma^{\mu\nu}$. One can see that the different initial conditions approach an attractor, which is not described by NS. This becomes more evident when we increase $\eta/s$ by a factor of two, shown in Fig.\ \ref{fig1} (right panel). We checked some fluid elements located in other parts of the fireball, finding similar results. Interestingly enough, the system crosses the NS result and remains in a far-from-equilibrium state until the end of our time evolution ($\sim 10$ fm). This also shows that $\pi^{\mu\nu}$ depends on the spatial gradients in a complicated manner once transverse flow is included (even for a conformal plasma). However, a detailed analysis still needs to be performed - we will return to this point in a separate publication. Nevertheless, we believe that the results presented in this work already strongly indicate that attractors can persist even in realistic heavy ion collisions.  

\section{Slow-roll expansion for a general flow}
\label{sec3}

The slow-roll expansion first appeared in the context of relativistic hydrodynamics in \cite{Heller:2015dha}, and its lowest order truncation was shown to provide a good approximation for the attractor of IS theory undergoing Bjorken flow \cite{Heller:2015dha,Denicol:2017lxn,Romatschke:2017vte,Strickland:2017kux}. An important feature of the slow-roll expansion in hydrodynamics, studied in detail in \cite{Denicol:2017lxn,Denicol:2018pak}, is that at a given order this expansion contains derivative terms of all orders. Also, the 1st order truncation of the slow-roll series, when expanded in gradients, provides results that match the gradient expansion computed at 2nd order \cite{Denicol:2018pak}. Therefore, it is reasonable to expect that the slow-roll series can be relevant when studying attractor behavior in heavy ion collisions. 

In this contribution we show how to perform the slow-roll series in conformal IS theory under general flow conditions (going beyond boost invariance and radial symmetry). This can be done as follows. We first define the dimensionless variable $\mathcal{X}^{\mu\nu}  = \pi^{\mu\nu}/(\varepsilon+P)$ ($\sim$ the inverse Reynolds number) and use the conservation laws together with (\ref{eq1}) to find the following exact equation for $\mathcal{X}^{\mu\nu}$
\begin{equation}
\tau_\pi D\mathcal{X}^{\langle\mu\nu\rangle} =  -\mathcal{X}^{\mu\nu}+\frac{2}{5}\tau_\pi \sigma^{\mu\nu}- \frac{4}{3}\tau_\pi  \mathcal{X}^{\mu\nu} \mathcal{X}^{\alpha\beta}\sigma_{\alpha\beta}.
\label{defineISshearnew}
\end{equation}
One can see \cite{Denicol:2017lxn,Denicol:2018pak} that the slow-roll series may be implemented assuming that the derivatives of the inverse Reynolds number are small in comparison to the other terms in (\ref{defineISshearnew}) - this defines the attractor regime. This behavior can be systematically investigated by including a small book-keeping (dimensionless) parameter $\epsilon$ on the left-hand side of (\ref{defineISshearnew}), i.e.,
\begin{equation}
\epsilon\,\tau_\pi D\mathcal{X}^{\langle\mu\nu\rangle} =  -\mathcal{X}^{\mu\nu}+\frac{2}{5}\tau_\pi \sigma^{\mu\nu}- \frac{4}{3}\tau_\pi  \mathcal{X}^{\mu\nu} \mathcal{X}^{\alpha\beta}\sigma_{\alpha\beta},
\end{equation}
while looking for a series solution where $\mathcal{X}^{\mu\nu}(x;\epsilon) = \sum_{n=0}^\infty \epsilon^n \mathcal{X}_n^{\mu\nu}(x)$. At a given order in the truncation, the approximation for $\mathcal{X}^{\mu\nu}$ is found by gathering the terms and setting $\epsilon \to 1$. In general, this procedure becomes highly nontrivial as it requires expressing $\mathcal{X}^{\mu\nu}$ as a functional of the hydrodynamic fields and all of its space-like derivatives. Here we explicit compute the first term of the expansion (using the full result for $\varepsilon$ and $u^\mu$), which leads to a resummed expression for the shear viscosity. This 0th order truncation gives 
\begin{equation}
-\mathcal{X}_0^{\mu\nu}+\frac{2}{5}\tau_\pi \sigma^{\mu\nu}- \frac{4}{3}\tau_\pi  \mathcal{X}_0^{\mu\nu} \mathcal{X}_0^{\alpha\beta}\sigma_{\alpha\beta}=0.
\label{0thSRshear}
\end{equation} 
After contracting this equation with $\sigma_{\mu\nu}$, it is easy to see that the general solution must be such that $\mathcal{X}_0^{\mu\nu} \sim \sigma^{\mu\nu}$. We write $\mathcal{X}_0^{\mu\nu} = \frac{2}{5}\tau_\pi \sigma^{\mu\nu} \mathcal{S}$ with the scalar quantity $\mathcal{S}$ being defined by $\frac{8}{15}\tau_\pi^2 \sigma_{\mu\nu}\sigma^{\mu\nu}\mathcal{S}^2 + \mathcal{S} -1 =0$, which is obtained from (\ref{0thSRshear}). The solution that recovers the NS result and is well-defined for all values of the gradients is 
\begin{equation}
\mathcal{S}(K_N)= \frac{15}{16K_N^2}\left(\sqrt{1+\frac{32}{15}K_N^2}-1\right)
\end{equation}
where we defined $K_N = \tau_\pi \sqrt{\sigma_{\mu\nu}\sigma^{\mu\nu}}$. Therefore, the 0th order truncation of the slow-roll series gives $\pi^{\mu\nu} \to 2\eta^R(K_N) \sigma^{\mu\nu}$ where $\eta^R(K_N) =\eta\, \mathcal{S}(K_N)$ is the resummed shear viscosity coefficient. One can see that  $\eta^R \leq \eta$ for all values of gradients while it behaves as $\eta^R \sim 1/K_N$ when $K_N \gg 1$. Therefore, differently than what happens in NS theory, in IS $\pi^{\mu\nu}$ remains finite even when $K_N \gg 1$ due to an effective resummed shear viscosity that is a nontrivial function of the gradients, being directly sensitive to the properties of the underlying flow. Higher order terms in the slow-roll expansion are expected to lead to resummed expressions for 2nd order transport coefficients - we shall explore this and other related questions in future work.

\section{Conclusions}
\label{sec:conclusions}
In this contribution we showed that Lagrangian schemes to solve the viscous hydrodynamic equations can be instrumental in the study of attractors in heavy ion collisions, going beyond previous studies that involved highly symmetrical flow profiles. Our numerical results indicate that attractors do appear in current heavy ion collision simulations. Further studies are needed to understand how the different parts of the medium evolve, especially the edges of the fireball. Effects from freezeout, nontrivial equation of state, and bulk viscosity can easily be systematically studied in this approach. More importantly, one can compare the fate of attractors in large and small systems. In fact, if attractors also appear in the latter, that would help explain why typical hydrodynamic signatures persist even when we shrink the quark-gluon plasma. Furthermore, in this work we showed how to perform the slow-roll expansion in conformal Israel-Stewart theory undergoing a general flow, which naturally gave rise to a resummed shear viscosity coefficient that is a nontrivial function of the gradients of the hydrodynamic fields.





\section*{Acknowledgements}
GSD thanks Funda\c c\~ao Carlos Chagas Filho de Amparo \`a Pesquisa do Estado do Rio de Janeiro (FAPERJ), grant number E26/202.747/2018. JN acknowledges partial support from Funda\c c\~ao de Amparo \`a Pesquisa do Estado de S\~ao Paulo (FAPESP), grant number 2017/05685-2.




\end{document}